\documentclass[conference]{IEEEtran}
\usepackage[utf8]{inputenc} % custom change

\ifCLASSINFOpdf
\else
\fi
\usepackage{url}
\usepackage[backend=bibtex,style=ieee,natbib=true,doi=false,isbn=false,url=false]{biblatex} %added
\AtEveryBibitem{%
	\clearfield{language}%
}

\usepackage{xspace}

\usepackage{graphicx}
\usepackage{csquotes}
\usepackage{amsmath}
\usepackage{amsthm}
\usepackage{amssymb}
\usepackage{mathtools}

\usepackage{thmtools}

 \theoremstyle{definition}

 \theoremstyle{plain}
 
 \newtheorem{proposition}{Proposition}

 \numberwithin{definition}{subsection}
 \numberwithin{remark}{subsection}
 \numberwithin{theorem}{subsection}
 \numberwithin{proposition}{subsection}
 \numberwithin{lemma}{subsection}
 \numberwithin{corollary}{subsection}

\usepackage{algorithm}
\usepackage{algorithmic}

\usepackage{booktabs}
\usepackage{multicol}
\usepackage{adjustbox}

\usepackage[disable]{todonotes}

\DeclarePairedDelimiter\abs{\lvert}{\rvert}%

\usepackage{array}
\newcolumntype{L}[1]{>{\raggedright\let\newline\\\arraybackslash\hspace{0pt}}m{#1}}
\newcolumntype{C}[1]{>{\centering\let\newline\\\arraybackslash\hspace{0pt}}m{#1}}
\newcolumntype{R}[1]{>{\raggedleft\let\newline\\\arraybackslash\hspace{0pt}}m{#1}}

\usepackage{hyperref}
\usepackage[capitalise]{cleveref}

\newcommand{\clust}{C}
\newcommand{\clustering}{\mathcal{C}}
\newcommand{\dataset}{\mathcal{X}}
\newcommand{\clustreps}{\mathcal{R}}
\newcommand{\rep}{R}
\newcommand{\freqset}{\mathcal{F}}
\newcommand{\maxfreqset}{\mathcal{M}}
\newcommand{\origamisample}{\mathcal{S}}
\newcommand{\randmaxfreqgraph}{S}
\newcommand{\freqnodes}{\text{N}}
\newcommand{\freqpath}{\text{P}}
\newcommand{\struclus}{\emph{StruClus}\xspace}
\newcommand{\minsup}{\mathit{minSup}}
\newcommand{\minsupf}{\text{minSup}}
\newcommand{\supp}{\text{sup}}
\newcommand{\bin}{\text{B}}% binomial distribution
\newcommand{\cov}{\text{cov}}
\newcommand{\hcov}{\text{aCov}}
\newcommand{\relcov}{\text{relCov}}

\newcommand{\mcs}{\text{mcs}}
\newcommand{\simmcsmax}{\text{sim}}
\newcommand{\simmcsmaxthres}{\mathit{sim}_{\min}}
\newcommand{\numsep}{\mathit{sim}_{\text{num}}}
\newcommand{\maxrep}{\clustreps_{\max}}
\newcommand{\hs}{\mathit{hs}}
\newcommand{\ls}{\mathit{ls}}
\newcommand{\hr}{\mathit{hr}}
\newcommand{\lr}{\mathit{lr}}
\newcommand{\proclus}{PROCLUS\xspace}
\newcommand{\kkmeans}{Kernel K-Means\xspace}

\newcommand\numberthis{\addtocounter{equation}{1}\tag{\theequation}}

\let\oldtodo\todo
\renewcommand{\todo}[1]{\oldtodo[inline]{#1}}

\newcommand{\optional}[1]{}

\begin{document}
\title{StruClus: Structural Clustering of Large-Scale Graph Databases}

\author{\IEEEauthorblockN{Till Schäfer}
\IEEEauthorblockA{TU Dortmund University\\
Dept. of Computer Science\\
44227 Dortmund, Germany\\
Email: till.schaefer@cs.tu-dortmund.de}
\and
\IEEEauthorblockN{Petra Mutzel}
\IEEEauthorblockA{TU Dortmund University\\
Dept. of Computer Science\\
44227 Dortmund, Germany\\
Email: petra.mutzel@cs.tu-dortmund.de}}

\maketitle

\begin{abstract}
We present a structural clustering algorithm for large-scale datasets of small labeled graphs, utilizing a frequent subgraph sampling strategy. A set of representatives provides an intuitive description of each cluster, supports the clustering process, and helps to interpret the clustering results.
The projection-based nature of the clustering approach allows us to bypass dimensionality and feature extraction problems that arise in the context of graph datasets reduced to pairwise distances or feature vectors. While achieving high quality and (human) interpretable clusterings, the runtime of the algorithm only grows linearly with the number of graphs. Furthermore, the approach is easy to parallelize and therefore suitable for very large datasets. Our extensive experimental evaluation on synthetic and real world datasets demonstrates the superiority of our approach over existing structural and subspace clustering algorithms, both, from a runtime and quality point of view.
\end{abstract}


\IEEEpeerreviewmaketitle

\section{Introduction}
Molecules, protein interaction networks, XML documents, social media interactions, and image segments all have in common that they can be modeled by labeled graphs. The ability to represent topological and semantic information causes graphs to be among the most versatile data structures in computer science. In the age of Big Data, huge amounts of graph data are collected and the demand to analyze them increases with its collection.
We focus on the special case of clustering large sets of small labeled graphs. Our main motivation stems from the need to cluster large-scale molecular databases for drug discovery, such as PubChem\footnote{\url{https://pubchem.ncbi.nlm.nih.gov/}}, ChEMBL\footnote{\url{https://www.ebi.ac.uk/chembl/}}, ChemDB\footnote{\url{http://cdb.ics.uci.edu/}} or synthetically constructed de-novo databases \cite{KUW+2010}, which contain up to a billion molecules. However, the presented approach is not limited to this use case.

Clustering techniques aim to find homogeneous subsets in a set of objects.
Classical approaches do not interpret the objects directly,
but abstract them by utilizing some intermediate representation, such as feature vectors or pairwise distances.
While the abstraction over pairwise distances is beneficial in terms of generality, it can be disadvantageous in the case of \emph{intrinsic high dimensional} datasets \cite{CN2001,GKL2003}. In this case the \emph{concentration effect} may cause the pairwise distances to loose their relative contrast; i.e. the distances converge towards a common value \cite{BGRS1999}. The concentration effect is closely related to a bad \emph{clusterability} \cite{AB2009}.

Sets of graphs are usually clustered by transforming the graphs to feature vectors or by using graph theoretic similarity measures.

Typical feature extraction methods for graphs are: counting \textit{graphlets}, that is, small subgraphs \cite{WWK2008,SVP+2009}, counting \textit{walks} \cite{VSKB2010}, and using \textit{eigenvectors} of adjacency matrices (spectral graph theory) \cite{FPV2014}. The enumeration of all subgraphs is considered intractable even for graphs of moderate size, because there exist up to exponentially many subgraphs wrt the graph size.
Many efficient clustering algorithms have been proposed for vector space. Therefore, the transformation to feature vectors might look beneficial in the first place. However, the above mentioned feature extraction methods tend to produce a large amount of distinct and often unrelated features. This results in datasets with a high intrinsic dimensionality \cite{TK2006,KMS2014a}.
Additionally, the extracted features only approximate the graph structure, which
implies that feature vectors cannot be transformed back into a graph. Hence, the interpretability of clustering algorithms which perform vector modifications (e.g. calculating centroids) is limited.

Besides the utilization of various feature extraction techniques, it is possible to compare graphs directly using graph theoretic distances such as \textit{(maximum) common subgraph} derived distances \cite{BS1998,WSKR2001,FV2001} or the \textit{graph edit distance} \cite{Bun1997}. The computation of the previously mentioned graph theoretic distances is NP-hard and as shown in \cite{KMS2014a} their application results in datasets with a high intrinsic dimensionality as well. High quality clustering methods for arbitrary (metric) and high dimensional datasets furthermore require a superlinear number of exact distance computations. These factors render graph theoretic distance measures in combination with generic clustering algorithms infeasible for large-scale datasets.

Subspace and projected clustering methods tackle high dimensional datasets by identifying subspaces in which well separated clusters exist. However, generic subspace algorithms come with a high runtime burden and are often limited to an euclidean vector space.

Our \textit{structural projection} clustering algorithm approaches the dimensionality problems by explicitly selecting cluster representatives in form of common subgraphs.
Consider a feature mapping to binary feature vectors that contain one feature for each subgraph that is found in the complete dataset. For each graph, its feature vector has binary entries encoding the presence of the associated substructure graph. Selecting a common subgraph $S$ as cluster representative is another way of selecting a subspace in these feature vectors consisting of all features associated with subgraphs of $S$.

Our main contributions in this paper are: We present a novel structural projection clustering algorithm for datasets of small labeled graphs which scales linearly with the dataset size. A set of representatives provides an intuitive description of each cluster, supports the clustering process, and helps to interpret the clustering results. Up to our knowledge, this is the first approach actively selecting representative sets for each cluster based on a new ranking function. The candidates for the representatives are constructed using frequent subgraph sampling.
In order to speed up the computation, we suggest a new error bounded sampling strategy for support counting in the context of frequent subgraph mining.
Our experimental evaluation shows that our new approach outperforms competitors in its runtime and quality.

The paper is structured as follows: \cref{sec:related} provides an overview of related clustering algorithms. Basic definitions are given in \cref{sec:prelim}. \cref{sec:struclus} presents the main algorithm and a runtime analysis.
Our experimental evaluation in which we compare our new algorithm to SCAP \cite{SKK2014}, PROCLUS \cite{APW+1999} and \kkmeans \cite{Gir2002} is presented in \cref{sec:evaluation}.

\section{Related Work}\label{sec:related}
Several clustering algorithms for graph and molecule data have been proposed in the last years. \citet{TK2006} presented an EM algorithm using a binomial mixture model over very high dimensional binary vectors, indicating the presence of all frequent substructures. Two years later, \citet{TK2008} presented a similar graph clustering algorithm using a Dirichlet Process mixture model and pruning the set of frequent substructures to achieve smaller feature vectors. A K-Median like graph clustering algorithm has been presented by \citet{FVS+2009}, that maps each graph into the euclidean space utilizing the edit distance to some pivot elements. A median graph is then selected based on the distance to the euclidean median. Furthermore, a parallel greedy overlapping clustering algorithm has been presented by \citet{SBSK2011}. It adds a graph to a cluster whenever a common substructure of a user-defined minimum size exists. However, none of the previously mentioned algorithms are suitable for large datasets as a result of their high computational complexity. XProj \cite{ATW+2007} uses a projection-based approach by selecting all (enumerated) frequent substructures of a fixed size as cluster representatives.
The approach scales well with the dataset size but is limited to trees. A generalization to graphs would result in a huge performance degradation.
Furthermore, there exist some hybrid approaches, that pre-cluster the dataset by using a vector-based representation and refine the results using structural clustering algorithms. The most relevant with respect to large-scale datasets is the SCAP algorithm proposed by \citet{SKK2014}.

Of course many subspace algorithms are also applicable after using a feature extraction method. Giving a comprehensive overview on subspace techniques is out of the scope of this article. However, there are two algorithms that are of special interest for this work: In \cite{YM2003} it is shown that frequent pattern mining can be used for feature selection in vector space. This relates to XProj and our approach in the way, that the selection of a graph representative---with the help of frequent substructure mining---is another way of selecting a subspace in the feature space of substructures. In the later evaluation we will compare ourself to the \proclus \cite{APW+1999} algorithm. It is a fast projected clustering algorithm with noise detection, that selects features by minimizing variance.
The algorithm has been studied intensively and performed well in various subspace clustering comparisons \cite{MGAS2009, PM2006}.

\section{Preliminaries}\label{sec:prelim}
An \emph{undirected labeled graph} $G=(V,E,l)$ consists of a finite set of \emph{vertices} $V(G)=V$, a finite set of \emph{edges} $E(G)=E \subseteq \{ \{u,v\} \subseteq V \mid u\not=v \}$ and a labeling function $l:V\biguplus E \rightarrow L$, where $L$ is a finite set of \emph{labels}.
$\abs{G}$ is used as a short term for $\abs{V(G)} + \abs{E(G)}$.
A \emph{path} of length $n$ is a sequence of vertices $(v_0, \dots, v_n)$ such that $\{v_i,v_{i+1}\} \in E$ and $v_i \not= v_j$ for $i \not= j$.
Let $G$ and $H$ be two undirected labeled graphs. A \emph{(label preserving) subgraph isomorphism} from $G$ to $H$ is an injection $\psi:V(G) \rightarrow V(H)$, where $\forall v \in V(G): l(v) = l(\psi(v))$ and $\forall u,v \in V(G) : \{u,v\} \in E(G) \Rightarrow \{\psi(u),\psi(v)\} \in E(H) \wedge l(\{u,v\}) = l(\{\psi(u),\psi(v)\})$. Iff there exists a subgraph isomorphism from $G$ to $H$ we say $G$ is \emph{supported} by $H$, $G$ is a \emph{subgraph} of $H$, $H$ is a \emph{supergraph} of $G$ or write $G \subseteq H$.
If there exists a subgraph isomorphism from $G$ to $H$ and from $H$ to $G$, the two graphs are isomorphic.
A \emph{common subgraph} of $G$ and $H$ is a graph $S$, that is subgraph isomorphic to $G$ and $H$.
Furthermore, the \emph{support} $\supp(G, \mathcal{G})$ of a graph $G$ over a set of graphs $\mathcal{G}$ is the fraction of graphs in $\mathcal{G}$, that support $G$. $G$ is said to be \emph{frequent}, iff its support is larger or equal than a \emph{minimum support threshold} $\minsup$. A frequent subgraph $G$ is \emph{maximal}, iff there exists no frequent supergraph of $G$. For a set of graphs $\mathcal{G}$, we write $\mathcal{\freqset(\mathcal{G})}$ for the set of all frequent subgraphs and $\mathcal{\maxfreqset(\mathcal{G})}$ for the set of all maximal frequent subgraphs.
A \emph{clustering} of a graph dataset $\dataset$ is a partition $\clustering = \{\clust_1, \dots, \clust_n\}$ of $\dataset$.  Each \emph{cluster} $\clust \in \clustering$ consists of a set of graphs and is linked to a \emph{set of cluster representatives} $\clustreps(\clust) =\{ \rep_1,\dots,\rep_k  \}$ which are itself undirected labeled graphs. Please note, that we consider each graph in our dataset to be a distinct object. As a result, it is possible to have isomorphic graphs in a single set.

\section{The \struclus Algorithm}\label{sec:struclus}
A high level description of the \struclus algorithm is given in \cref{fig:mainAlgorithm}. Initially, it partitions the dataset using a lightweight pre-clustering algorithm. Afterwards, the clustering is refined using an optimization loop similar to the K-Means algorithm. In order to fit the number of clusters to the dataset structure (i.e., to achieve a good cluster separation and homogeneity, see \cref{sec:repupdate,sec:splitmerge}), we use a cluster splitting and merging strategy in each iteration.

\begin{algorithm}
	\caption{\struclus Algorithm}
	\label{fig:mainAlgorithm}
	\begin{algorithmic}[1]
		%     \FUNCTION{StruClus}{}
		\STATE apply pre-clustering \COMMENT{\cref{sec:preclustering}}
		\WHILE{not convergent \COMMENT{\cref{sec:convergence}}}
		\STATE split clusters \COMMENT{\cref{sec:splitmerge}}
		\STATE merge clusters \COMMENT{\cref{sec:splitmerge}}
		\STATE update representatives \COMMENT{\cref{sec:repupdate}}
		\STATE assign graphs to closest cluster \COMMENT{\cref{sec:assignment}}
		\ENDWHILE
		%     \ENDFUNCTION
	\end{algorithmic}
\end{algorithm}

An important ingredient of our algorithm is the set of representatives $\clustreps(\clust)$ for each cluster $\clust \in \clustering$.
Representatives serve as an intuitive description of the cluster and define the substructures over which intra cluster similarity is measured.
The set $\clustreps(\clust)$ is chosen such that for every graph $G \in \clust$ there exists at least one representative $R \in \clustreps(\clust)$ which is subgraph isomorphic (i.e., supported by $G$).
With the exception of a single \emph{noise} cluster the following invariant holds after each iteration:
\begin{equation}\label{eq:clusterinv}
\forall \clust \in \clustering : \forall G \in \clust : \exists R \in \clustreps(\clust) : R \subseteq G.
\end{equation}

\begin{figure*}[!t]
	\centering
	\raisebox{-.5\height}{%
		\includegraphics[width=0.96\columnwidth]{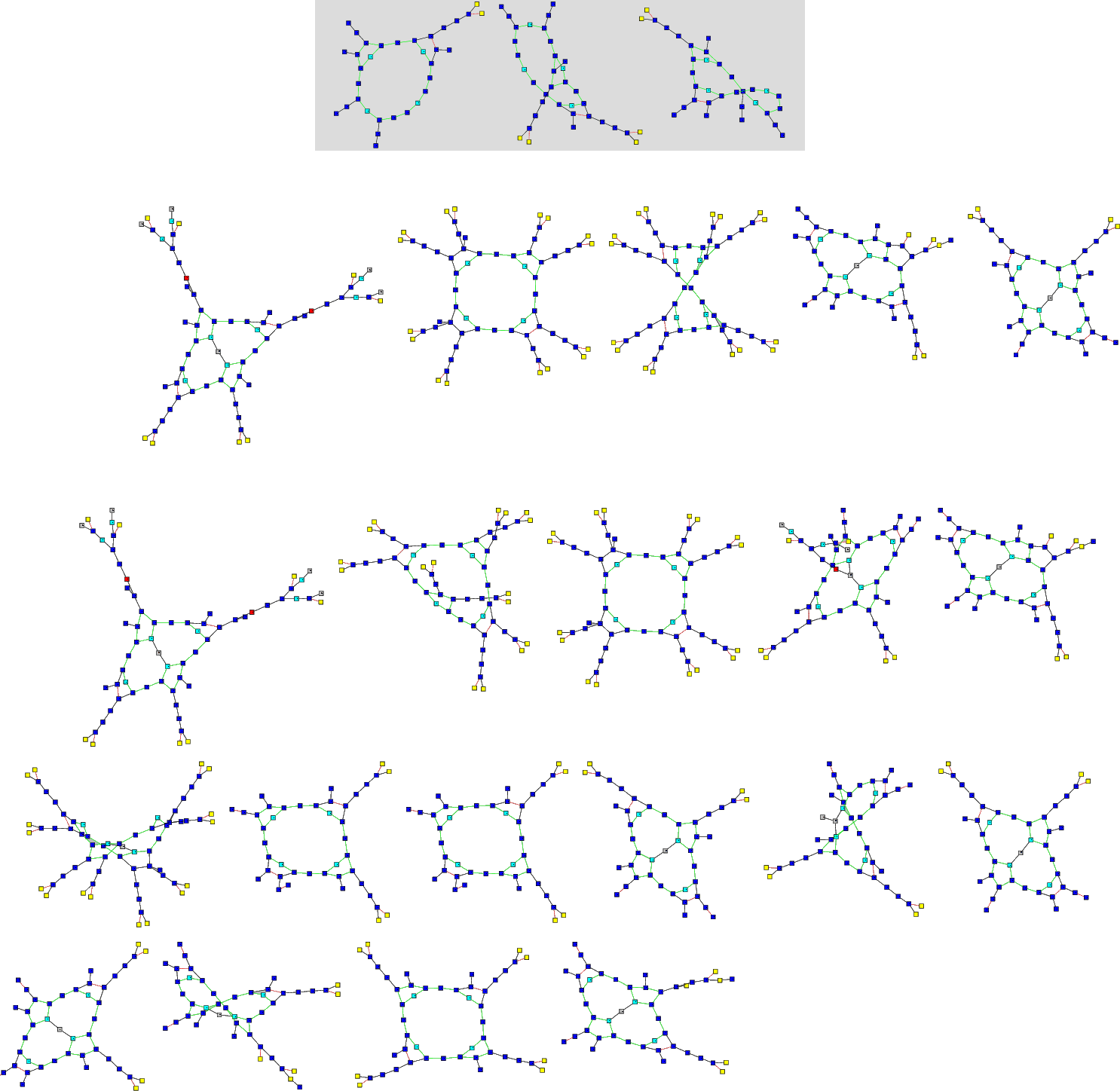}
	}%
	\qquad
	\raisebox{-.5\height}{%
		\includegraphics[width=0.96\columnwidth]{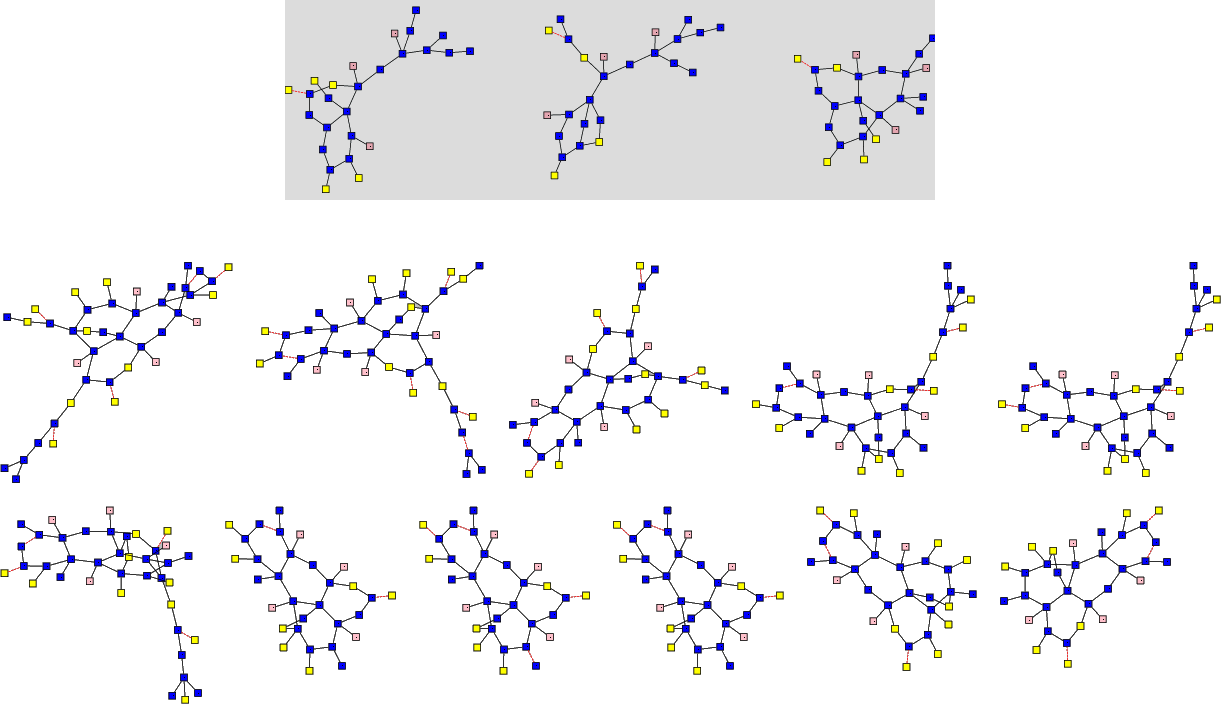}
	}
	\caption{Two real world clusters generated with \struclus. The grey boxes show the cluster representatives.}
	\label{fig:rwclusters}
\end{figure*}

The representative set $\clustreps(\clust)$ of a cluster $\clust \in \clustering$ is constructed using maximal frequent subgraphs %$\maxfreqset(\clust)$. 
of $\clust$ (see \cref{sec:repmining}). Having a representative set instead of a single representative has the advantage, that graphs composed of multiple common substructures can be represented. In order to be meaningful and human interpretable, the cardinality of $\clustreps(\clust)$ is limited by a user defined value $\maxrep$. \Cref{fig:rwclusters} shows two example clusters and their representatives of a real world molecular dataset generated with \struclus.

\subsection{Stochastic Representative Mining}\label{sec:repmining}
We construct the representative set $\clustreps(\clust)$ of a cluster $\clust \in \clustering$ using maximal frequent subgraphs 
of $\clust$. Since the set $\maxfreqset(\clust)$ may have exponential size wrt the
maximal graph size in $\clust$, we restrict ourselves to a subset of
candidate representatives $\origamisample(\clust) \subseteq \maxfreqset(\clust)$
using a randomized maximal frequent connected subgraph sampling technique from ORIGAMI \cite{HCS+2007}
combined with a new stochastic sampling strategy for support counting.
In a second step, the final representative set $\clustreps(\clust) \subseteq \origamisample(\clust)$ is selected using a ranking function (see \cref{sec:repselection}).

ORIGAMI constructs a maximal frequent connected subgraph $\randmaxfreqgraph \in \origamisample(\mathcal{G})$ over a set of graphs $\mathcal{G}$ by extending a random frequent vertex with frequent paths of length one leading to a graph $S'$.
% In ORIGAMI a maximal frequent connected subgraph $\randmaxfreqgraph \in \origamisample(\clust)$ is constructed by extending a random frequent node with frequent paths of length one.
In a first step, all frequent vertices $\freqnodes(\mathcal{G})$ and all frequent paths of length one $\freqpath(\mathcal{G})$ are enumerated with a single scan of $\mathcal{G}$. Then, for each extension, a random vertex of $S'$ is chosen and a random, label preserving path $p \in \freqpath(\mathcal{G})$ is connected to it in a forward (creating a new vertex) or backward (connecting two existing vertices) fashion. After each extension, the support $\supp(S',\mathcal{G})$ is evaluated by
solving a subgraph isomorphism test for all graphs in $\mathcal{G}$. If $\supp(S',\mathcal{G}) \geq \minsup$, the extension is permanently added to $S'$ or otherwise removed. If no further extension is possible without violating the minimum support threshold, a maximal frequent subgraph $\randmaxfreqgraph$ has been found. This process is justified by the \emph{monotonicity property} of subgraphs $\mathcal{G}_\subseteq$ of graphs in $\mathcal{G}$: 
\begin{align*}%
%  \mathcal{G}_\subseteq &:= \{ G ~\vert~ G \subseteq H \in \mathcal{G} \}\\
\forall G,H \in \mathcal{G}_\subseteq &: G \subseteq H \Rightarrow \supp(G, \mathcal{G}) \geq \supp(H, \mathcal{G}) \numberthis \label{eq:monotonicity}
\end{align*}
While ORIGAMI greatly improves performance in comparison with enumeration algorithms, the $\abs{\mathcal{G}}$ subgraph isomorphism tests for each extension remain a major performance bottleneck for \struclus. For this reason, we have added a stochastic sampling strategy for support counting. 

Initially, we draw a random sample $\mathcal{H} \subseteq \mathcal{G}$. Then $\hat{\theta} = \supp(S',\mathcal{H})$ is an estimator for the parameter $\theta$ of a binomial distribution $\bin(\cdot, \theta)$, where $\theta = \supp(S',\mathcal{G})$ is the true probability of the underlying Bernoulli distribution. 
We are interested if the true value of $\supp(S',\mathcal{G})$ is smaller than the minimum support threshold. 
Without loss of generality, let us focus on the case $\hat{\theta} < \minsup$ in the following. 
We can take advantage of a binomial test under the null hypothesis, that $\theta \geq \minsup$ and thereby determine the probability of an error, if we assume $\supp(S',\mathcal{G}) < \minsup$. With a predefined significance level $\alpha$ we can decide if the sample gives us enough confidence to justify our assumption. If we cannot discard our null hypothesis, we continue by doubling the sample size $\abs{\mathcal{H}}$ and repeat the process. In the extreme case we will therefore calculate the exact value of $\supp(S', \mathcal{G})$. \optional{The initial and minimal sample size $\abs{\mathcal{H}_{\min}}$ is set to $30$ to prevent testing problems with small sample sizes.}

The statistical test is repeated for each extension and each sample size doubling. As a consequence, a multiple hypothesis testing correction is necessary to bound the real error for $\randmaxfreqgraph$ to be a maximal frequent substructure of $\mathcal{G}$.

\begin{proposition}\label{prop:hyptestcorr}%
	Let $\mathcal{G}$ be a set of undirected labeled graphs, $\abs{\mathcal{H}_{\min}}$ the minimal sample size, 
	$\freqpath(\mathcal{G})$ the set of all frequent paths of length one, 
	$\minsup$ a minimum support threshold, and $V_{\max}$ the  $(1 - \minsup)$-quantile of the sorted (increasing order) graph sizes of each graph in $\mathcal{G}$. Then the maximal number of binomial tests to construct a maximal frequent substructure over $\mathcal{G}$ is bounded by:
	\begin{equation*}
	\left\lceil\log\frac{\abs{\mathcal{G}}}{\abs{\mathcal{H}_{\min}}}\right\rceil ~ (V_{\max}^2 + V_{\max}) ~ \abs{\freqpath(\mathcal{G})}
	\end{equation*}
\end{proposition}
\begin{proof}
	The sample size is doubled $\left\lceil\log\frac{\abs{\mathcal{G}}}{\abs{\mathcal{H}_{\min}}}\right\rceil$ times if the test never reaches the desired significance level.
	The size of some $\randmaxfreqgraph \in \maxfreqset(\mathcal{G})$ is bounded by the size of each supporting graph. In the worst case $\randmaxfreqgraph$ is supported by the ($\abs{\mathcal{G}}  \minsup$)-largest graphs of $\mathcal{G}$. The graph size of the smallest supporting graph is then equal to the $(1 - \minsup)$-quantile of the sorted graph sizes in increasing order.
	The number of backward extensions is bounded by the number of vertex pairs times the number of applicable extensions $p \in \freqpath(\mathcal{G})$. Additionally, we need to check $\abs{\freqpath(\mathcal{G})}$ forward extensions for each vertex to conclude that $\randmaxfreqgraph$ is maximal.
\end{proof}

Finally, with the value of \cref{prop:hyptestcorr} we are able to apply a Bonferroni correction to our significance level. We can afford a relative high error, because the selection of the final representatives will filter out bad candidates. However, the used significance level has an influence on the runtime. On the one hand a high error leads to many bad candidates and we need to increase the number of candidates to mine. On the other hand a low error will lead to larger sample sizes to reject the null hypothesis. A maximal error of $50\%$ has turned out to be a good choice during our experimental evaluation.

\subsection{Update Representatives}\label{sec:repupdate}
\subsubsection{Representative Selection}\label{sec:repselection}
In its role as a cluster description, a good representative $\rep \in \clustreps(\clust)$ explains a large portion of its cluster. Accordingly, it should (a) be supported by a large fraction of $\clust$ and (b) cover a large fraction of vertices and edges of each graph $G \in \clust$ supporting $R$. We define the coverage of a graph $G$ by a representative $\rep$ as $\cov(\rep,G) := \frac{\abs{\rep}}{\abs{G}}$.
The two criteria are closely related to the cluster homogeneity. A uniform cluster, that is, a cluster that contains only isomorphic graphs, can achieve optimal values for both criteria.
Vice versa, the monotonicity property \eqref{eq:monotonicity} implies non-optimal values for inhomogeneous clusters for at least one of the two criteria.
As homogeneous clusters are desired, we use a product of the two criteria for our ranking function.

In order to discriminate clusters from each other, a cluster representative should be cluster specific as well. Thus, its support in the rest of the dataset should be low.
For this reason, we use the following ranking function for a dataset $\dataset$, cluster $\clust$ and representative $\rep \in \clustreps(\clust)$:
\begin{align*}
\clust_{\rep} &:= \{G \in \clust ~\vert~ \rep \subseteq G\}\\
\text{rank}(\rep) &:= \frac{\abs{\clust_{\rep}} ~ \abs{\rep}}{\sum_{G \in \clust_{\rep}} \abs{G}} ~ (\supp(\rep, C) - \supp(\rep, \dataset))\numberthis\label{eq:repranking}
\end{align*}
Finally, we select the $\maxrep$ highest ranked sampled subgraphs from $\origamisample(\clust)$ as cluster representatives $\clustreps(\clust)$.

\subsubsection{Balancing Cluster Homogeneity}\label{sec:balancehomo}
Besides the representative selection, the choice of the minimum support threshold for representative mining has an influence on the cluster homogeneity. The fraction of unsupported graphs for a cluster $\clust$ after updating the cluster representatives is bounded by ($1-\minsup$). The bound clearly relates to criteria (a) from the representative ranking (see \cref{sec:repselection}). However, unsupported graphs are removed from the cluster and will be assigned to a different cluster at the end of the current iteration (see \cref{sec:assignment}). Due to the monotonicity property \eqref{eq:monotonicity}, this process of sorting out graphs 
by choosing a minimum support threshold below $1$, 
will increase the size of the representatives and therefore our coverage value, i.e criteria (b). A decrease of the minimum support threshold will lead to an increase in the size of the representatives. Subsequently, this process will also reduce the cluster cardinality. Therefore, increasing the homogeneity to an optimal value will result in a clustering with uniform or singleton clusters, which is clearly not the desired behavior.

To get around this, we will aim towards a similar homogeneity for all clusters and choose the minimum support threshold cluster specific. However, choosing a fixed homogeneity level a priori is not an easy task, as an appropriate value depends on the dataset. Therefore we will calculate an average coverage score over all clusters and use this as a baseline adjustment. For the ease of computation, we choose a slightly simplified coverage approximation:
\begin{align}
\hcov(\clust) &= \frac{\frac{1}{\abs{\clustreps(\clust)}} \sum_{R \in \clustreps(\clust)} \abs{R}}{\frac{1}{\abs{\clust}} \sum_{G \in \clust} \abs{G}}\\
\relcov(\clust, \clustering) &= \frac{\hcov(\clust)}{\frac{1}{\abs{\clustering}}\sum_{\clust' \in \clustering}\hcov(\clust')}
\end{align}
Finally, we can define a linear mapping from the relative coverage $\relcov$ to a cluster specific minimum support threshold with the help of two predefined tuples $(\ls, \lr)$ and $(\hs, \hr)$, where $\ls < \hs$ and $\lr < \hr$. The parameter $\ls$ ($\hs$) denotes the lowest (highest) support value and $\lr$ ($\hr$) the relative coverage value mapped to the lowest (highest) minimum support:
\begin{align*}
\text{f}(\clust, \clustering) &:=\relcov(\clust, \clustering) ~ \frac{\hs - \ls}{\hr - \lr} + (\ls - \lr ~ \frac{\hs - \ls}{\hr - \lr})\\
\minsupf(\clust, \clustering) &:=  \begin{cases}
\ls & \text{if } \relcov(\clust, \clustering) < \lr\\
\hs & \text{if } \relcov(\clust, \clustering) > \hr\\
\text{f}(\clust, \clustering) & \text{otherwise.}
\end{cases}\numberthis
\end{align*}
To result in a minimum support threshold of $1$ for all clusters (i.e. stopping the process of sorting out graphs if the clustering is balanced), we will set the values of the parameters $\hs$ very close or equal to $1$ and $\hr<1$.

\subsection{Cluster Assignment}\label{sec:assignment}

Each graph $G$ in the dataset is assigned to its \emph{most similar cluster} in the assignment phase. As a measure for similarity, we are summing up the squared sizes of the representatives of a cluster, which are subgraph isomorphic to $G$. This choice of similarity is once more justified by the representative ranking criteria. We square the representative sizes, to prefer a high coverage over a high number of representatives to be subgraph isomorphic to the assigned graph. \optional{If a cluster is empty after the assignment phase, it will be removed from the clustering.}

As mentioned in \cref{sec:balancehomo} it is possible that a graph $G \in \clust$ will no more be supported by any representative $\rep \in \clustreps(\clust)$ of its cluster $\clust$ after updating the representatives.
In this situation we will create a single noise cluster, where all graphs (of all clusters) that are not supported by any representative  are collected. As the minimum support threshold is bounded by the fixed value $\ls$ (see \cref{sec:balancehomo}) and the number of representatives is limited by $\maxrep$, it is not guaranteed that we can find an appropriate set of representatives for this most likely largely inhomogeneous noise cluster. It is therefore excluded from the invariant \eqref{eq:clusterinv}.

The problem of finding all subgraph isomorphic graphs in a graph database is also known as the \emph{subgraph search problem} and was extensively studied in the past. We apply the fingerprint pre-filtering technique CT-Index \cite{KKM2011}, which has emerged from this research topic, to speed up the assignment phase. CT-Index enumerates trees and circles up to a specified size for a given graph and hashes the presence of these subgraphs into a binary fingerprint of fixed length. If the fingerprint of a graph $G$ has a bit set that is unset in the fingerprint of a graph $H$ we can conclude that no subgraph isomorphism from $G$ to $H$ exists, because $G$ contains a subgraph that is not present in $H$. We calculate a fingerprint for each graph and representative and only perform a subgraph isomorphism test in our assignment phase if the fingerprint comparison cannot rule out the presence of a subgraph isomorphism.

\subsection{Cluster Splitting and Merging}\label{sec:splitmerge}

Without the operation of cluster splitting, 
the above mentioned process of creating noise clusters would create at most one extra cluster in each iteration of our main loop. 
A large difference in the initial and final number of clusters therefore would lead to a slow convergence of the \struclus algorithm towards its final result. As mentioned before, it is also possible that no representative is found at all for the noise cluster, and therefore the process of sorting out graphs from the noise cluster to increase its homogeneity is stopped completely in the worst case.
A similar situation can occur for regular clusters. For example, if a cluster is composed of uniform sets $T \in \mathcal{T}$ of graphs, we will require a minimum support threshold less than or equal to $1 - \frac{1}{\min_{T \in \mathcal{T}} \{\abs{T}\}}$ to sort the smallest possible number of graphs out.
For this reason, a cluster splitting step is necessary (see \cref{fig:mainAlgorithm}). In this step, all clusters that have a relative coverage value below an a priori specified threshold $\relcov_{\min}$ will be merged into a single set of graphs and the pre-clustering algorithm is applied on them. The resulting clusters are added back to the clustering.

On the contrary to cluster splitting, which focuses on cluster homogeneity, cluster merging ensures a minimum separation between clusters. Separation can be measured on different levels. Many classical measures define separation as the minimum distance between two cluster elements. However, this type of definition is not suitable for projected clustering algorithms, because the comparison does not take the cluster specific subspace into account. As mentioned in the introduction, the cluster representatives in \struclus define the subspace of the cluster. Additionally, they serve as a description of the graphs inside the cluster itself.
A high coverage value leads to an accurate cluster description. Thus, we will define \emph{separation} between two clusters $\clust$ and $\clust'$ solely over the representatives sets $\clustreps(\clust)$ and $\clustreps(\clust')$. This definition is also beneficial from a runtime perspective, as separation calculation is independent of the cluster size.
Without cluster merging it is possible that clusters with very similar representatives do exist. Although the pre-clustering will ensure that the initial clusters will have dissimilar representatives (see \cref{sec:preclustering}) it may happen that two clusters converge towards each other or that newly formed clusters are similar to an already existing one. Therefore, we will merge two clusters whenever their representatives are too similar.

To compare two single representatives $\rep,\rep'$ we calculate the size of their maximum common subgraph (MCS) and use its relative size as similarity:
\begin{align}\label{eq:simmcsmax}
\simmcsmax(\rep, \rep') := \frac{\abs{\mcs(\rep, \rep')}}{\max\{\abs{\rep}, \abs{\rep'}\}}
\end{align}
The maximum of the representatives sizes is chosen as denominator to support different clusters with subgraph isomorphic representatives, which differ largely in size. Finally, we will merge two clusters $\clust$ and $\clust'$ if the following condition holds:
\begin{equation}
\begin{aligned}
\abs{\{(\rep, \rep') \in \clustreps(\clust) \times \rep(\clust') ~ \vert ~ \simmcsmax(\rep, \rep')\\
	\geq \simmcsmaxthres \}} \geq \numsep
\end{aligned}
\end{equation}
where $\numsep$ is a minimum number of representative pairs which have a similarity greater than or equal to $\simmcsmaxthres$. Note that the calculated MCS between two representatives $\rep \in \clustreps(\clust), \rep' \in \clustreps(\clust')$ is supported by all the graphs $G \in \clust \cup \clust'$, that support either $\rep$ or $\rep'$, because the subgraph isomorphism relation is transitive. The coverage for these graphs in the merged cluster is furthermore bounded by $\max\{\cov(G,\rep), \cov(G,\rep')\} ~ \simmcsmax(\rep, \rep')$ if we reuse the MCS as representative. For this reason, we recommend to set $\numsep$ close to the number of representatives per cluster to support a large fraction of graphs in the merged cluster.
The parameter $\simmcsmaxthres$ is furthermore an intuitive knob to adjust the granularity of the clustering.

Finally, the representatives for the merged clusters are updated. We do not use the calculated MCSs as representatives, because better representatives may exist and the decision problem \enquote{Does an MCS larger than some threshold exist?} is computationally less demanding than calculating the MCS itself.

\subsection{Pre-Clustering}\label{sec:preclustering}

The pre-clustering serves as an initial partitioning of the dataset. A random partitioning of all graphs would be problematic as representatives may not be found for all partitions and the found representatives are most likely not cluster specific. This will result in a high number of clusters to be merged to a few inhomogeneous clusters and in a slow convergence of the \struclus algorithm.

To pre-cluster the dataset $\dataset$, we compute maximal frequent subgraphs $\origamisample(\dataset) \subseteq \maxfreqset(\dataset)$, as described in \cref{sec:repmining} with a fixed minimum support. These frequent subgraphs serve as representative candidates for the initial clusters. To avoid very similar representatives we will first greedily construct maximal sets of dissimilar graphs. As a measure of similarity we re-use the similarity \eqref{eq:simmcsmax} and the threshold $\simmcsmaxthres$ from cluster merging. In other words, we are picking all graphs $G$ from $\origamisample(\dataset)$ in a random order and add $G$ to our dissimilar set $\mathcal{D}$, if $\nexists H \in \mathcal{D}$ with $\simmcsmax(G,H) < \simmcsmaxthres$. This process is repeated several times and the largest set $\mathcal{D}_{\max}$ is used to create one cluster for each $H \in \mathcal{D}$ with $H$ as single representative. Afterwards we run a regular assignment phase as described in \cref{sec:assignment}.
As a result of re-using $\simmcsmaxthres$ we can expect, that we have well separated cluster and no cluster merging is necessary in the first iteration of the main loop (excluding the noise cluster).

\subsection{Convergence}\label{sec:convergence}

\struclus optimizes cluster homogeneity while maintaining a minimal cluster separation. As described in \cref{sec:balancehomo}, homogeneity is defined by two criteria wrt the cluster representatives. With an exception of the noise cluster, the representative support criteria (a) can be dismissed, because not represented graphs will be sorted out of the cluster. It was further discussed, that an optimal homogeneity can be achieved for singleton clusters. However, the later introduced cluster separation constraint will limit the granularity of the clustering, because two similar representatives will be merged. 
Nevertheless, there might exist several clusterings with different granularity that respect the separation constraint. For this reason the objective function balances the coverage criteria (b) of the homogeneity and the granularity of the clustering. The parameter $\alpha$ adjusts this granularity:
\begin{align}
\text{z}(\clustering) &:= \frac{\sum_{\clust \in \clustering} \abs{\clust} ~ \hcov(\clust)}{\abs{\clustering}^\alpha}
\end{align}
As a consequence of the cluster splitting and merging, the objective function will fluctuate and contain local optima. We will therefore smooth the objective function. Let $\clustering_i$ be the clustering after the $i$-th iteration. \cref{fig:mainAlgorithm} will terminate after the first iteration $c$ for which the following condition holds:
\begin{align}
c \geq s + w \wedge \frac{\sum_{c-w < i \leq c}\text{z}(\clustering_i)}{\sum_{c-s-w < i \leq c-s}\text{z}(\clustering_i)} \leq 1 + \beta
\end{align}
where $w$ is the averaging width and $\beta$ is the minimum relative increase of the objective function in $s$ iterations.

\subsection{Runtime Analysis}
The subgraph isomorphism problem and the maximum common subgraph problem are both NP-complete even in their decision variants \cite{GJ1979}. \struclus solves these problems in order to calculate the support and to decide if clusters need to be merged.
% Furthermore, the size of a representative is bounded by the sizes of the supporting graphs.
Furthermore, the size of a representative can be linear in the sizes of the supporting graphs.
Thus, \struclus scales exponentially wrt the size of the graphs in the dataset $\dataset$. Nevertheless, these problems can be solved sufficiently fast for small graphs with a few hundred vertices and edges, e.g., molecular structures. We will therefore consider the graph size a constant in the following analysis and focus on the scalability wrt the dataset size. Let $\mathcal{V}_{\max}$ be the maximal number of vertices for a graph in $\dataset$ and $\clustering_{\max}$ the maximal number of clusters during the clustering process.

\paragraph*{Representative Mining}{
As described in \cref{prop:hyptestcorr} the number of extensions to mine a single maximal frequent subgraph from a set of graphs $\mathcal{G} \subseteq \dataset$ is bounded by $(V_{\max}^2 + V_{\max}) ~ \abs{\freqpath(\mathcal{G})}$. The variable $V_{\max}$ is obviously a constant. Also $\abs{\freqpath(\mathcal{G})}$ is only bounded by $\mathcal{V}_{\max}$ and the lowest minimum support threshold $\ls$, as a fraction of $\frac{\lceil \ls ~ \abs{\mathcal{G}} \rceil}{\mathcal{V}_{\max}^2}$ of all edges in $\mathcal{G}$ must be isomorphic to be frequent. Thus, there exist at most $\frac{\mathcal{V}_{\max}^2}{\ls}$ frequent paths of length one. Note, that the number of distinct labels does not have an influence on this theoretical bound. For each mined maximal frequent subgraph we need to calculate the support over $\mathcal{G}$ and this involves $\mathcal{O}(\abs{\mathcal{G}})$ many subgraph isomorphism tests. Therefore, the runtime to mine a single maximal frequent subgraph is bounded by $\mathcal{O}(\abs{\mathcal{G}})$.
}

\paragraph*{Representative Update}{
Updating the representatives involves two steps: representative mining and representative selection. Representative mining includes the calculation of the cluster specific minimum support with the help of $\relcov$. The $\relcov$ values for each cluster can be calculated in $\mathcal{O}(\clustering_{\max})$ if we maintain a sum of the graph sizes of each cluster and update this value during cluster assignment. Since the cluster sizes sum up to $\abs{\dataset}$ the runtime of the actual frequent subgraph mining is in $\mathcal{O}(\dataset)$. To rank a representative candidate $\rep \in \clustreps(\clust)$, the set $\clust_\rep$, the value $\sup(\rep,\clust)$, and the value $\sup(\rep,\dataset)$ can be calculated by a single scan over $\dataset$, i.e. calculating whether $\rep$ is subgraph isomorphic to each graph in $\dataset$. As the number of candidates per cluster is a constant, the runtime of the ranking is in $\mathcal{O}(\clustering_{\max}~\abs{\dataset})$, which is also the overall runtime of the representative update.
}

\paragraph*{Other Parts}{
The runtimes of the cluster assignment and the pre-clustering are in $\mathcal{O}(\clustering_{\max} ~ \abs{\dataset})$. Cluster splitting is computable in  $\mathcal{O}(\clustering_{\max} + \abs{\dataset})$, cluster merging in $\mathcal{O}(\clustering_{\max}^2 + \abs{\dataset})$, and converge in $\mathcal{O}(\clustering_{\max})$ time.
}

\paragraph*{Overall Runtime}{
	A single iteration of \cref{fig:mainAlgorithm} has a runtime of $\mathcal{O}(\clustering_{\max} ~ \abs{\dataset})$, and hence is linear. This is justified by the observation that $\mathcal{O}(\clustering_{\max}^2) \leq \mathcal{O}(\clustering_{\max} ~ \abs{\dataset})$.
	Let $m$ be the number of iterations after \cref{fig:mainAlgorithm} terminates. Then, the overall runtime is in $\mathcal{O}(m~\clustering_{\max} ~ \abs{\dataset})$.
}

\section{Evaluation}\label{sec:evaluation}
In the following evaluation we compare \struclus to SCAP \cite{SKK2014}, \proclus \cite{APW+1999} and \kkmeans \cite{Gir2002} wrt their runtime and the clustering quality. Furthermore, we evaluate the influence of our sampling strategy for support counting and evaluate the parallel scaling of our implementation.

\paragraph*{Hardware \& Software}{
All tests were performed on a dual socket NUMA system (Intel Xeon \optional{CPUs }E5-2640 v3\optional{ @ 2.60GHz, 8 cores + HT}) with 128 GiB of RAM. The applications were pinned to a single NUMA domain (i.e. 8 cores + HT / 64 GiB RAM) to eliminate random memory effects. Turbo Boost was deactivated to minimize external runtime influences.
Ubuntu Linux 14.04.1 was used as operating system. The Java implementations of \struclus, \proclus and \kkmeans were running in an Oracle Java Hotspot VM 1.8.0\_66. SCAP was compiled with GCC 4.9.3 and -O3 optimization level. \struclus and SCAP are shared memory parallelized.
}

\paragraph*{Test Setup \& Evaluation Measures}{
The tests were repeated 30 times if the runtime was below 2 hours and 15 times otherwise. Quality is measured by ground truth comparisons. We use the Normalized Variation of Information (NVI) \cite{Mei2007} and Fowlkes-Mallows (FW) \cite{FM1983} measures for \struclus, \proclus and \kkmeans. While NVI and FW are established quality measures, they are not suited for overlapping clusterings as produces by SCAP. Thus, Purity \cite{dataclustering-chap-validation} was used for comparison with SCAP.
}

\begin{table*}[t!]
	\centering
	\caption{Real world dataset statistics. Cumulated values are given as triple min / max / average.}
	\label{tab:rwdatastat}
	\begin{adjustbox}{max width=\textwidth}
		\begin{tabular}{rrlllll}
			\toprule[1.5pt]
			dataset & size & classes & \# vertices & \# edges & \# vertex labels & \# edge labels\\
			\midrule\\
			AnchorQuery & $65\,700$ & $11$ & $11$ / $90$ / $79.19$ & $11$ / $99$ / $86.02$ & $6$ & $5$ \\
			Heterocyclic & $10\,000$ & $39$ & $9$ / $69$ / $42.99$ & $10$ / $79$ / $47.35$ & $25$ & $5$\\
			ChemDB & $5\,000\,000$ & -- & $1$ / $684$ / $50.74$ & $0$ / $745$ / $53.20$ & $86$ & $5$\\
			\bottomrule[1.5pt]
		\end{tabular}
	\end{adjustbox}
\end{table*}

\paragraph*{Datasets}{
We evaluate the algorithms on synthetic datasets of different sizes and three real world datasets. The synthetic datasets have $\approx35$ vertices and  $\approx51$ edges on average with $10$ vertex and $3$ edge labels (weights drawn from an exponential distribution). They contain $100$ clusters and $5\%$ random noise graphs. For each cluster $3$ seed patterns were randomly generated with a Poisson distributed number of vertices (mean $10$) and an edge probability of $10\%$. Additionally, we created a common seed pool with $100$ (noise) seeds. The cluster specific graphs were then generated by connecting the cluster specific seeds with $0$ to $2$ common noise seeds.
The first real world dataset (AnchorQuery) is a molecular de-novo database. Each molecule is the result of a chemical reaction of multiple purchasable building blocks. We have used $11$ reaction types, i.e., class labels, from AnchorQuery\footnote{\scriptsize{\url{http://anchorquery.csb.pitt.edu/reactions/}}}. 
The second real world dataset (Heterocyclic) is similar to the first one, but contains heterocyclic compounds and $39$ distinct reaction types. 
The third real world dataset (ChemDB) contains $5$ million molecules from ChemDB \cite{chemdb2007}. ChemDB is a collection of purchasable molecules from $150$ chemical vendors. The dataset has no ground truth and will mainly serve as proof that we are able to cluster large real world datasets. 
\Cref{tab:rwdatastat} shows additional statistics about these real world datasets.

To cluster the datasets with \proclus the graph data was transformed to vectors by counting distinct subgraphs of size $3$, resulting in $7\,000$ to $10\,000$ features on the synthetic datasets. The application to the AnchorQuery and Heterocyclic datasets results in much lower feature counts of $274$ and $133$. The same vectors were used as features space representation for \kkmeans. Additionally, we have evaluated various other graph kernels with explicit feature mapping for \proclus and \kkmeans, such as the Weisfeiler-Lehman shortest path, Weisfeiler-Lehman subtree and fixed length random walk kernels \cite{SSvL+2011}. However, we observed feature vectors with even higher dimensionality and clustering results with lower quality. For this reason, the results of these graph kernels are omitted here.
}

\paragraph*{Algorithm Configuration}{
The parameters of \struclus were set as follows: maximal error for support counting: $50\%$; number of representative candidates: $\optional{\abs{\origamisample(\cdot)} = }25$; minimum support threshold calculated with: $\lr = 0; \ls = 0.4; \hr = 0.9; \hs = 0.99$; splitting threshold: $\hcov \leq 0.6$; minimum separation $\simmcsmaxthres = 0.3; \numsep = 3$; convergence determined with: $\alpha = 1, \beta=0.01, w=3, s=3$.
SCAP has two parameters: (a) the minimum size for a common substructure, that must be present in a single cluster and (b) a parameter that influences the granularity of the fingerprint-based pre-partitioning. (a) was set to $80\%$ of the mean seed size for the synthetic dataset, otherwise to $8$. (b) was set to the highest number that results in a reasonable clustering quality, as the clustering process is faster with fine grained pre-partitioning, but cannot find any clusters over the pre-partition boundary. For the synthetic dataset this was $0.2$.
\proclus has two parameters as well: (a) the number of clusters and (b) the average dimensionality of the cluster subspaces. (a) was set to the number of clusters in the ground truth. The ChemDB dataset, which has no ground truth, was too large for \proclus. (b) was set to $20$, for which we got best quality results. The number of clusters is the only parameter of \kkmeans and was set in exact the same manner as for \proclus. Please note, that the a priory selection of this optimal value gives \proclus and \kkmeans an advantage over their competitors during the evaluation.
}

\begin{table*}[t!]
	\caption{Results for the synthetic datasets. Runtimes are given in hours. Best results are marked bold. CV is the coefficient of variation. The CV value is given as the maximum for each column. All other values are averaged. Quality measures are: Normalized Variation of Information (NVI), Fowlkes-Mallows Index (FW), and Purity.}
	\label{tab:synthetic}
	\begin{adjustbox}{max width=\textwidth}
		\begin{tabular}{rlllllllllllllll}
			\toprule[1.5pt]
			\multicolumn{1}{c}{\textbf{Size}} & \multicolumn{4}{c}{\textbf{\struclus}} & \multicolumn{3}{c}{\textbf{\struclus (Exact Support)}} & \multicolumn{2}{c}{\textbf{SCAP}} & \multicolumn{3}{c}{\textbf{\proclus}} & \multicolumn{3}{c}{\textbf{\kkmeans}}\\
			                  & Runtime          & NVI             & FM              & Purity          & Runtime & NVI             & FM              & Runtime                     & Purity & Runtime & NVI    & FM       & Runtime & NVI    & FM     \\\\
			CV\hspace{-8pt}   & $<0.08$          & $<0.03$         & $<0.07$         & $<0.03$         & $<0.07$ & $<0.02$         & $<0.04$         & $<0.06$                     & $<0.02$& $<0.32$ & $<0.11$& $<0.22$  & $<0.01$ & $<0.01$ & $<0.01$ \\
			\cmidrule(l){2-5}                                                        \cmidrule(l){6-8}                             \cmidrule(l){9-10}                     \cmidrule(l){11-13}           \cmidrule(l){14-16} 
			$1\,000$          & $0.05$           & $\mathbf{0.90}$ & $0.75$          & $\mathbf{0.90}$ & $0.05$  & $\mathbf{0.90}$ & $\mathbf{0.77}$ & $\mathbf{7.5\times10^{-4}}$ & $0.83$ & $0.13$  & $0.58$ & $0.26$   & $0.02$  & $0.77$ & $0.54$ \\
			$5\,000$          & --               & --              & --              & --              & --      & --              & --              & --                          & --     & $4.99$  & $0.50$ & $0.24$   & $2.87$  & $0.84$ & $0.67$ \\
			$10\,000$         & $0.19$           & $\mathbf{0.95}$ & $\mathbf{0.87}$ & $\mathbf{0.99}$ & $0.59$  & $0.93$          & $0.85$          & $\mathbf{0.03}$             & $0.83$ & $11.91$ & $0.49$ & $0.24$   & $10.32$ & $0.86$ & $0.78$ \\
			$50\,000$         & $\mathbf{0.33}$  & $\mathbf{0.94}$ & $\mathbf{0.87}$ & $\mathbf{0.99}$ & $2.69$  & $0.92$          & $0.85$          & $0.38$                      & $0.83$ & --      & --     & --       & --      & --     & --    \\
			$100\,000$        & $\mathbf{0.47}$  & $0.93$          & $0.86$          & $\mathbf{0.99}$ & --      & --              & --              & $1.21$                      & $0.86$ & --      & --     & --       & --      & --     & --    \\
			$500\,000$        & $\mathbf{1.35}$  & $0.93$          & $0.86$          & $\mathbf{0.99}$ & --      & --              & --              & $18.15$                     & $0.83$ & --      & --     & --       & --      & --     & --    \\
			$1\,000\,000$     & $\mathbf{2.73}$  & $0.91$          & $0.84$          & $0.98$          & --      & --              & --              & --                          & --     & --      & --     & --       & --      & --     & --    \\
			\bottomrule[1.5pt]
		\end{tabular}
	\end{adjustbox}
\end{table*}

\paragraph*{Results}{
As shown in \cref{tab:synthetic}, \struclus outperforms all competitors in terms of quality on the synthetic datasets. 
The differences of all corresponding mean quality values are always larger than small multiples of the standard deviations. For small synthetic datasets, SCAP was the fastest algorithm. This changes for larger data set sizes. \struclus outperforms SCAP by a factor of $\approx13$ at size $500\,000$. Furthermore, we were unable to cluster the largest dataset with SCAP in less than $2$ days. \proclus and \kkmeans were the slowest algorithms with a huge gap to their next competitor. The sublinear growth of \struclus's runtime for the smaller datasets sizes is caused by our sampling strategy for support counting. Running \struclus with exact support counting yields a linear growth in runtime. It is important to mention that the clustering quality is not significantly influenced by the support counting strategy.

Our implementation of \struclus scales well with the number of cores as shown in \cref{fig:corescaling}. With $8$ cores we get a speedup of $7.15$. Including the Hyper-Threading cores, we get a speedup of $9.11$.
\begin{figure}[!ht]
	\centering
	\includegraphics[width=\columnwidth]{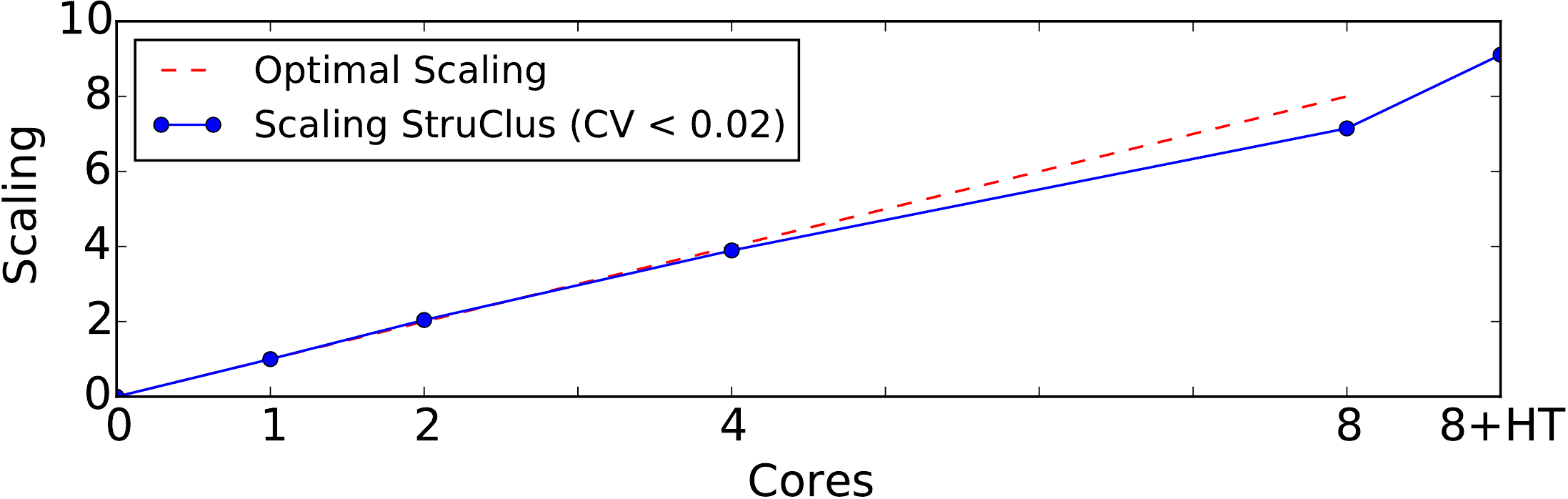}
	\caption{Parallel scaling: synthetic dataset of size $10\,000$}
	\label{fig:corescaling}
\end{figure}

\begin{table*}[t!]
	\centering
	\caption{Results for the real world datasets. Runtimes are given in hours. Best results are marked bold. CV is the coefficient of variation. The CV value is given as the maximum for each column. All other values are averaged. The ChemDB measurement was repeated only 3 times. We have not calculated the ---otherwise meaningless--- CV for it. Results for \kkmeans are are given for a random subset of the Heterocylcic dataset. Quality measures are: Normalized Variation of Information (NVI), Fowlkes-Mallows Index (FW), and Purity.}
	\label{tab:realworld}
	\begin{adjustbox}{max width=\textwidth}
		\begin{tabular}{rllllllllllll}
			\toprule[1.5pt]
			\multicolumn{1}{c}{\textbf{dataset}} & \multicolumn{4}{c}{\textbf{\struclus}} & \multicolumn{2}{c}{\textbf{SCAP}} & \multicolumn{3}{c}{\textbf{\proclus}} & \multicolumn{3}{c}{\textbf{\kkmeans}}\\
			                  & Runtime          & NVI             & FM              & Purity          & Runtime                     & Purity & Runtime          & NVI    & FM       & Runtime & NVI    & FM     \\\\
			CV\hspace{-8pt}   & $<2.91$          & $<0.09$         & $<0.12$         & $<0.08$         & $<0.02$                     & $<0.01$& $<0.03$          & $<0.05$& $<0.08$  & $<0.01$     & $<0.01$    & $<0.01$ \\
			                  \cmidrule(l){2-5}                                                        \cmidrule(l){6-7}                      \cmidrule(l){8-10}            \cmidrule(l){11-13} 
			AnchorQuery       & $\mathbf{2.47}$  & $0.44$          & $0.63$          & $0.89$          & --                          & --     & --               & --     & --       & --      & --     & --     \\
			Heterocyclic      & $1.07$           & $\mathbf{0.46}$ & $\mathbf{0.53}$ & $\mathbf{0.66}$ & $\mathbf{0.01}$             & $0.58$ & $\mathbf{0.01}$  & $0.29$ & $0.29$   & $\mathit{3.03}$ (Subset)      & $\textit{0.27}$     & $\mathit{0.29}$      \\
			ChemDB            & $\mathbf{\approx19}$ & --          & --              & --              & --                          & --     & --               & --     & --       & --      & --     & --     \\
			\bottomrule[1.5pt]
		\end{tabular}
	\end{adjustbox}
\end{table*}

The evaluation results wrt the real world datasets are summarized in \cref{tab:realworld}. We were unable to cluster the AnchorQuery and the ChemDB datasets with \proclus, \kkmeans and SCAP (even for high values of parameter (b)) in less than $2$ days. For the chemical reaction datasets, \struclus also needs more time compared to a synthetic dataset of similar size. This runtime increase can be partially explained by the larger graph sizes. However, other hidden parameters of the datasets, such as the size of maximal common substructures have an influence on the runtime as well. \struclus had some runtime outliers on the Heterocyclic dataset. They were caused by small temporary clusters with large ($>60$ vertices) representatives. The high runtime of the SCAP algorithm on the AnchorQuery dataset is a bit surprising, as the common substructures processed by SCAP are limited in their size (maximum $8$ vertices). We consider a larger frequent pattern search space to be the reason for this runtime increase. \kkmeans was surprisingly slow on the Heterocyclic dataset and took more than $24$ hours for a single run. We have therefore created a random subset with a size of $5000$ graphs for it. 

\struclus always outperforms the competitors wrt the quality scores. For the AnchorQuery dataset \struclus created $26$ clusters on average. The high score for the Purity measure shows, that \struclus splitted some of the real clusters, but keeps a well inter cluster separation. ChemDB was clustered by \struclus in $\approx19$ hours with $\approx 117$ clusters. As a consequence of the high runtime of the ChemDB measurement and the lack of competitors, we repeated the test only 3 times. The $\hcov$ value for the final clustering was $0.49$ on average. This highlights the ability of \struclus to cluster large-scale real-world datasets.
}

\section{Conclusion}
In this paper we have presented a new structural clustering algorithm for large scale datasets of small labeled graphs. With the help of explicitly selected cluster representatives, we were able to achieve a linear runtime in the worst case wrt the dataset size. A novel support counting sampling strategy with multiple hypothesis testing correction was able to accelerate the algorithm significantly without lowering the clustering quality. We have furthermore shown, that cluster homogeneity can be balanced with a dynamic minimum support strategy for representative mining. A cluster merging and splitting step was introduced to achieve a well separated clustering even in the high dimensional pattern space. Our experimental evaluation has shown that our new approach outperforms the competitors wrt clustering quality, while attaining significantly lower runtimes for large scale datasets. 
Although we have shown that \struclus greatly improves the clustering performance compared to its competitors, de-novo datasets with several billion of molecules are still outside the scope of this work. For this reason, we consider the development of a distributed variant of the algorithm to be future work. Another consideration to further improve the quality of the algorithm is to integrate a discriminative frequent subgraph miner for representative mining. The integration of the discriminative property into the mining process has the advantage, that higher quality representative candidates are mined. This will result in a lower number of necessary candidate patterns, which has a positive effect on the runtime. Furthermore, it allows to mine highly discriminant, non-maximal subgraphs. However, it is non-trivial to extend the support counting sampling strategy to such miners. Additionally, discriminative scores are usually non-monotonic on the subgraph lattice \cite{YCHY2008,TCG+2010}, which imposes another runtime burden.


\section*{Acknowledgment}
This work was supported by the German Research Foundation (DFG), priority programme \emph{Algorithms for Big Data (SPP 1736)}. We would like to thank Nils Kriege for providing a fast subgraph isomorphism and Madeleine Seeland, Andreas Karwath, and Stefan Kramer for providing their SCAP implementation.



% \bibliographystyle{IEEEtran}
% \bibliography{literatur.bib}

\printbibliography %added

\end{document}